# Brewster angle as never seen before

Alejandro Doval and Raúl de la Fuente

Nanomateriais, Fotónica e Materia Branda, Departamento de Física Aplicada, Universidade de Santiago de Compostela, Spain.

In this paper we will discuss a demonstration we have been performing for years; not only with physics students from our university (Universidade de Santiago de Compostela), but also with high school students in some talks aimed at encouraging them to study science. It is related to Brewster's angle and its visualization in an ingenious way using a "loaded" LCD monitor. In some way, this experiment is a reverse version of what happens when a vampire faces a mirror and sees no reflected image of himself.

**Context**

When studying light reflection at the interface between two transparent media[i] (see the BOX), there are two angles showing some interesting singular properties. The first one is the critical angle, the smallest angle of incidence that yields total reflection, provided that the incident medium is denser than the transmission one. The other angle having a particular relevance is the Brewster or polarizing angle, at which the reflected wave is linearly polarized perpendicularly to the plane of incidence. This happens regardless of the polarization state of the incident light: if its electric field vibrates inside the plane of incidence, the amplitude of the reflected wave will be null. The Brewster angle is easily found know the refraction indexes. Concretely, whether the incident medium is air, we will obtain the Brewster law:

$$tan\theta_B = n \qquad (1)$$

Being $n$ the refraction index of the second material. For instance, for a typical value such as $n = 1.5$, the corresponding Brewster angle is $\theta_B \approx 56.30°$. Some straightforward ways to measure the polarizing angle can be found in literature[ii,iii,iv,v].

## BOX. Electromagnetic theory of reflection and refraction

When an electromagnetic wave arrives at an interface between two media with different optical properties, a fraction of the incident radiation is refracted towards the second medium, while the rest is reflected backwards to the original one. It is convenient to split this problem into two simpler ones in order to study it, which differ in the polarization of the wave. In the first of them, the electric field of the incident wave points and oscillates in some direction included in the incidence plane (which is defined by the propagation vector of the incident wave and the direction normal to the boundary surface). The second problem involves a wave whose electric field is polarized perpendicular to the incidence plane.

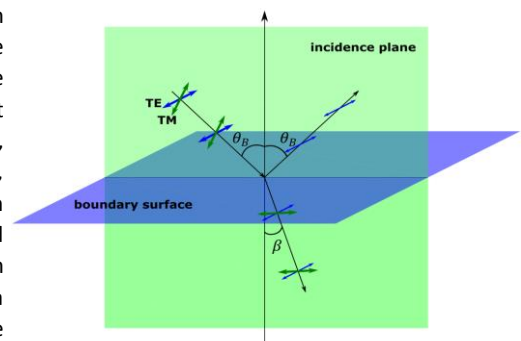

If the incident angle equals Brewster's angle, the reflected component whose electric field is parallel to the incidence plane is cancelled. In this case, the only reflected wave left is the one which has it electric field normal to the incident plane (provided this component is non-zero for the incident wave). The surface therefore acts as a polarizing filter by reflection, being its polarization axis normal to the incidence plane. This Brewster condition is depicted below.

The inherent properties of this angle are useful for many practical applications, mainly related to reflection elimination. It can help a fisherman wearing polarized sunglasses to get a better view of the river bed, or a driver to avoid being blinded by light. It is also practical to improve the quality of the pictures in a catalogue, or to increase photographic contrast when a polarizer filter is used in photography. However, these effects are not obvious and their explanation to the student can be arduous. Here we propose a simple experiment that allows a quick visualization of the characteristic properties of the polarizing angle, as well as the changes in the polarization of light when being reflected or transmitted at a surface.

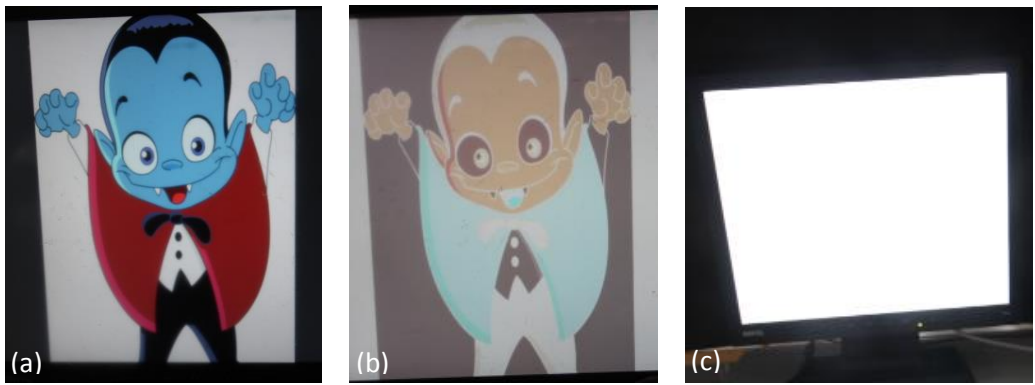

Fig. 1 The protagonists of our experiment: Blue-Vamp, Brown-Vamp and "secret" screen.

Before starting our experiment, we shall introduce its protagonists: Blue-Vamp and Brown-Vamp. We can see both in Fig. 1, projected on a colour LCD[vi] screen. The colours of the vampire in Fig. 1(b) are complementary to those in Fig. 1(a). There is a third main character playing a crucial role in our experience: a 'secret'[vii] LCD screen (Fig. 1(c)), whose front polarizing filter[viii] has been previously removed. It is convenient to add that the axis of this subtracted polarizer is vertical. When the monitor is on, it is completely lit emitting white light, regardless of the molecular orientation of its liquid crystals; and therefore, irrespective of the image we would see on a normal screen. In order to observe the projected image, we need a polarizing filter which reveals the different polarization states of each pixel, allowing the observation of the desired colours thanks to the RGB filters that are part of our LCD.

For this purpose, a suitable option is to benefit from the existence of Brewster's angle, using a transparent plate oriented in such a way that observation satisfies the Brewster condition. Doing this, we face an impressing phenomenon: a white screen with a projected colour image associated with reflection. Its explanation is rather simple if we know how the LCD screen works: reflection at the Brewster angle plays the role of the removed polarizing filter.

**Experiments**

Fig. 2 shows the 'secret' screen together with two methacrylate plates. The first plate is oriented vertically at the right side in front of the screen, while the other one is placed horizontally below it. Looking at both of them we meet our vamp-protagonists again. We have a white light source and two images, but… images of what? The answer is, on one hand in the projection method of the LCD screen, and on the other, in the characteristics of the reflection process. We need to keep in mind that our 'secret' screen lacks its front polarizing filter, which allowed us to see coloured images. Its function is being performed by the methacrylate plates, due to the fact of the incidence angle being near to Brewster.

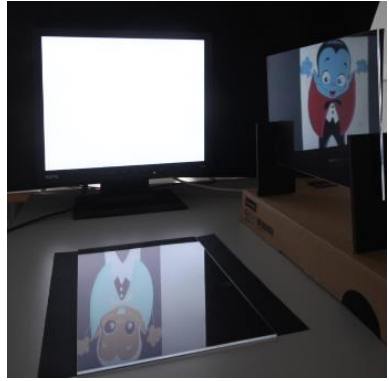

Fig. 2. Images observed using Brewster reflection of the screen.

Why are the pictures complementary? Because of how the plates are oriented. In the vertical one, the incidence plane is horizontal, and the only reflected light provided the Brewster condition is vertically polarized. In contrast, the incidence plane for the horizontal plate is vertical, and therefore, reflection corresponds to the horizontally polarized component of the original signal. If we put the polarizing sheet back to its original place on the screen correctly oriented (let us remember, that is with its polarization axes in the vertical direction), we would see the picture of Blue-Vamp on the screen (as in Fig. 1(a)), as well as its reflection at the vertical plate. And if we rotated the polarizer 90°, Brown-Vamp would appear on the screen (as in Fig. 1(b)), just like the corresponding reflection at the methacrylate plate laying on the table. Why is this last picture fainter than the previous one? Because the monitor tries to project intense and vivid images, making the complementary colours duller.

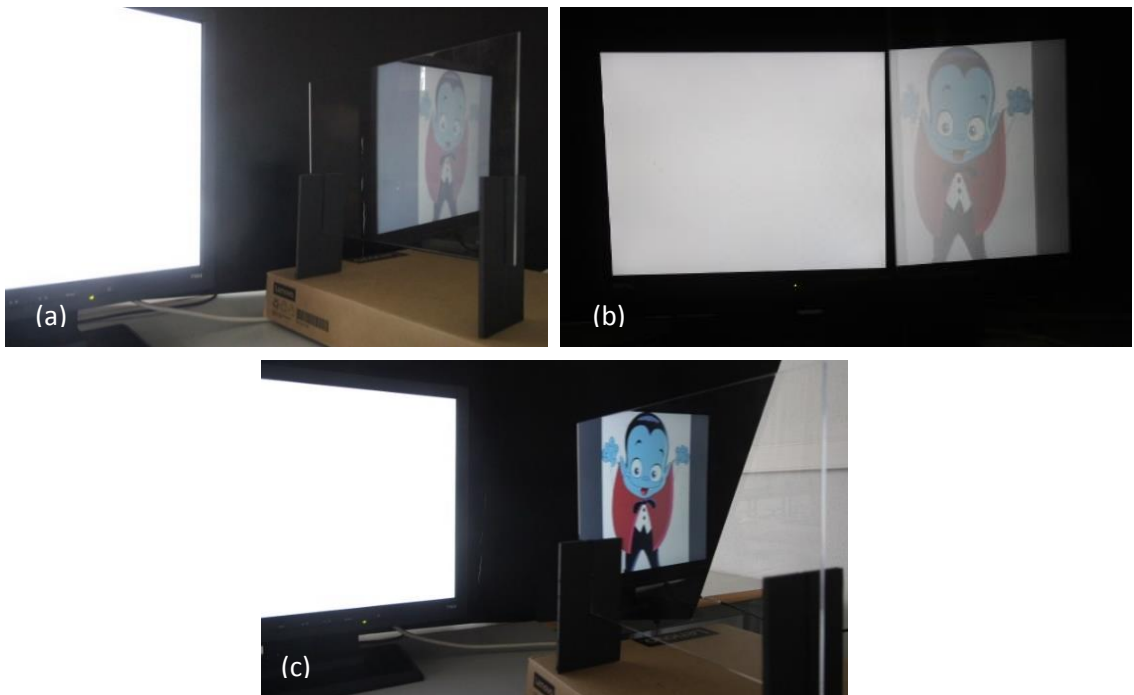

Fig 3. Reflected Blue-Vamp with incident angles below, over and around the Brewster angle.

Continuing our demonstration, we can perform another experience in which the position of reflector plates is varied, reaching observation angles far away from the Brewster condition. In that case, the images shown by the reflection are weaker in intensity. In Fig. 3, the results of locating the vertical plate at three different positions: near the screen (Fig. 3(a)), far away from it (Fig. 3(b)) and somewhere intermediate between those points (Fig.3 (c)). In the first case, the incidence angles are below Brewster, whereas in the second one they are over its value. The third photograph displays a more contrasted vampire, since the corresponding incidence angles are around the polarizing angle.

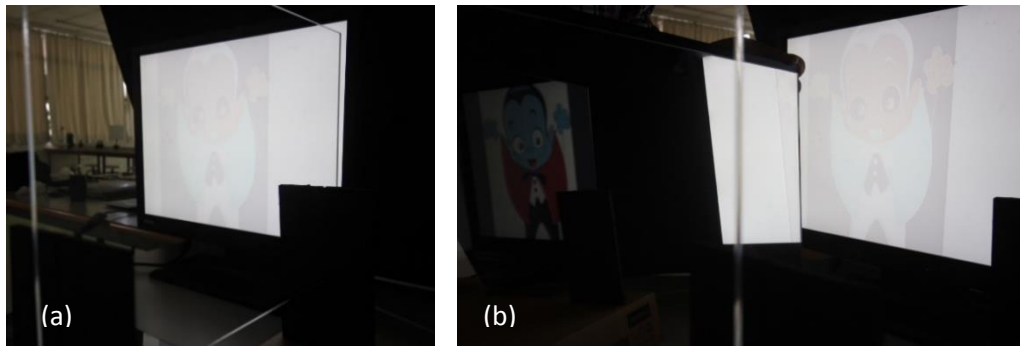

Fig. 4 Images observed using the transmitted light.

Let us now execute another idea. What happens if we look at the screen through one of the plates instead of the reflected light? In such case we will be able to watch an image complementary to that seen by reflection, as shown in Fig. 4(a). This is just a proof of energy conservation law. The incident light at a transparent media is divided between transmitted and reflected light, so that if both are summed together the original light is recovered (in our case, white light). Fig. 4(b) is the outcome of combining some imagination and a mirror. It is possible to see the two images at the same time: the one corresponding to the transmitted light (a faint Brown-Vamp) and the one corresponding to the reflected part, which is observed thanks to a retroreflection in a mirror (Blue-Vamp).

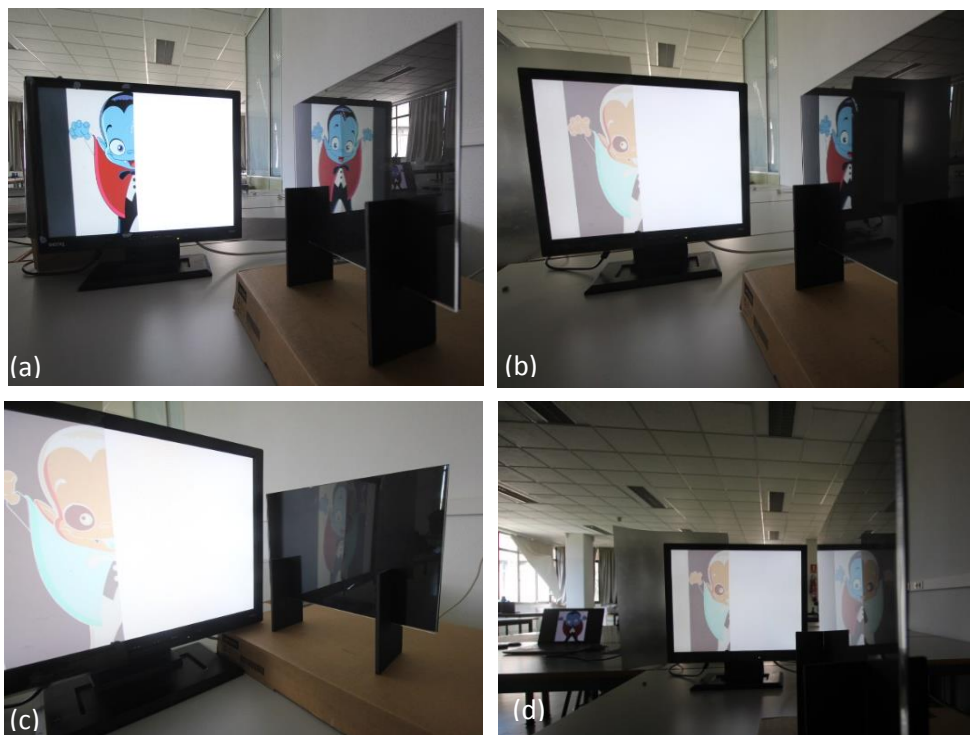

Fig. 5 Images taken with the polarizing filter covering the left half of the screen. In the first picture the axis of the polarizer is horizontal, whereas in the other three it is vertical. The upper photographs satisfy Brewster's condition, while the ones at the bottom show the observation from angles below (left) and above (right) Brewster.

A last experiment is suggested. We place the removed polarizer on the screen, but covering only a half of the vampire, changing the transmission axis trying its horizontal and vertical positions (Fig. 5). While explaining the shown images, lateral inversion in reflection (left to right and viceversa) must be taken into account. When the axis of the polarizer is horizontal the whole image can be seen, but the right side looks darker: it corresponds to the left side on the screen, which is covered by the polarizer, that absorbs some of the emitted light. In the rest of the displayed images, the axis of the polarizer on the screen is changed to the vertical position. For instance, the second one shows Brewster's configuration, and only the left part of the image can be seen (corresponding to the white background on the right of the screen). The other half, the image corresponding to the left side where Brown-Vamp can be seen directly, is completely black because of the axis of the polarizer on the screen being perpendicular to the reflection at the methacrylate plate. Furthermore, if we leave Brewster's configuration (just like in Fig. 3) towards greater or smaller angles, a faint image of Brown-Vamp starts to appear at the dark side, combining one half of each vampire.

Finally, let us show a last and impressive image! Yes, it is Blue-Vamp about to sleep (Fig. 6). Good night, dear friend! Hum..., sorry, good day! (vampires sleep during the day)

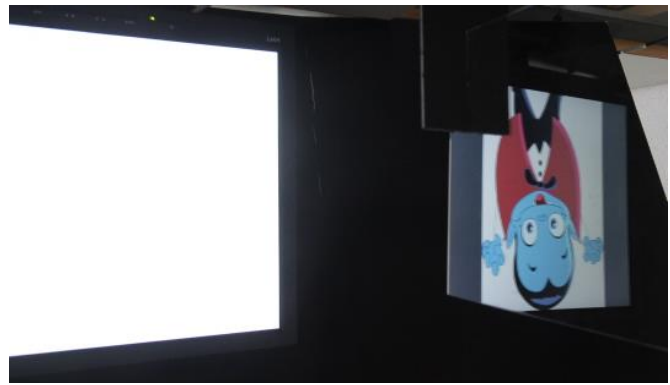

Fig. 6. Please, turn off the lights.

**Conclusions**

To sum up, it has been shown how polarizing filters work, together with an explanation of Brewster's configuration. Energy conservation at a surface between two different media has also been checked qualitatively: summing the reflected and transmitted energies we obtain the total incident amount. At the same time, we have also learnt that vampires are not so wicked as they are said to be. These experiences must be preceded by a qualitative explanation of the underlying physics of polarizing filters. It is also desirable that students have an idea of how an LCD screen works.

**Acknowledgements**

First author would like to thank the University of Santiago de Compostela and Banco de Santander for the granting of a scholarship under the programme "Bolsas de Iniciación á investigación para alumnos de máster" (Research Initiation Grants for Master's students).